\documentclass{anabs}
\usepackage{graphicx}
\usepackage{times}
\Volume{1-2}              
\Year{2003}              
\Month{02}               
\Pagespan{000}{000}      

\begin{document}
\lhead[\thepage]{F.E. Bauer: Title}
\rhead[Astron. Nachr./AN~{\bf 324} (2003) 1/2]{\thepage}
\headnote{Astron. Nachr./AN {\bf 324} (2003) 1/2, 000--000}

\title{The AGN Source Population in the Chandra Deep Field-North 
Survey: Constraints from X-ray Spectroscopy and Variability}

\author{F.~E.~Bauer,\inst{1}
C.~Vignali,\inst{1} D.~M.~Alexander,\inst{1} W.~N.~Brandt,\inst{1}
G.~P.~Garmire,\inst{1} A.~E.~Hornschemeier,\inst{1} P.~Broos,\inst{1}
L.~Townsley,\inst{1} and D.~P.~Schneider\inst{1} }
\institute{$^{1}$ Department of Astronomy \& Astrophysics, 525 Davey Lab, 
The Pennsylvania State University, University Park, PA 16802.}

\correspondence{fbauer@astro.psu.edu}

\maketitle

\section{Introduction}
\vspace{-0.11in}

With the 0.5-10.0~keV background nearly resolved (e.g., Giacconi et
al. 2000; Cowie et al. 2002), the emphasis has shifted towards
understanding the nature of the faint X-ray population --- a clear
necessity if we wish to understand the formation and evolution of
active galactic nuclei (AGN). Here we report on preliminary spectral
and temporal X-ray studies of AGN in the 2~Ms Chandra Deep Field-North
Survey (CDF-N; D.M. Alexander et al. in prep). Our sample is comprised
of 136 CDF-N sources with $>$~200 net counts ($\sim$30\% of the total
CDF-N sample), spanning $(0.014$--$2)\times10^{-13}$
ergs cm$^{-2}$ s$^{-1}$; 72 sources have known redshifts.

\vspace{-0.12in}
\section{Spectra}
\vspace{-0.11in}

We initially fitted each source with a fixed Galactic $N_{\rm H}$
($1.6\times10^{20}$~cm$^{-2}$; Stark et al. 1992), a variable
intrinsic $N_{\rm H}$ at the source redshift if known, and a simple
\hbox{powerlaw} with variable photon index $\Gamma$. Approximately 70\% of
the sources fitted by this model have $\chi^{2}_{\nu}<1.3$, with the
remaining sources often exhibiting soft excess residuals (i.e., due to
a partial covering or thermal component). The median spectral
parameters are $\Gamma=1.61$ and $N_{\rm H}=1.7\times10^{21}$
cm$^{-2}$ for all 136 sources, or $\Gamma=1.55$ and $N_{\rm
H}=2.6\times10^{21}$ cm$^{-2}$ for the 61 sources with $>$~500 counts. 
The intrinsic $N_{\rm H}$ distribution for the sample and a comparison
of $L_{\rm X}$ vs. $N_{\rm H}$ are shown in Fig.~1. The $N_{\rm H}$
distribution is strongly skewed towards low/Galactic values with only
a small fraction of sources lying above $10^{23}$ cm$^{-2}$. The 27
optically identified broad-line AGN (BLAGN) trace the overall sample.

\vspace{-0.12in}
\section{Emission Lines}
\vspace{-0.11in}

Ten of the 136 AGN ($\approx$7\%) exhibit obvious Fe K$\alpha$
emission-line features, with equivalent widths (EWs) of 0.1--1.3~keV. 
Two of the emission-line sources appear to be Compton thick AGN,
displaying both large EWs and extremely flat spectral slopes $\Gamma <
1.0$ characteristic of pure reflection (e.g., Maiolino et al. 1998). 
We have also constrained the number of potential Compton thick sources
among the 72 sources with known redshifts. By adding a Gaussian
component in the source spectrum ($E=6.4$~keV, $\sigma=0$~keV), we can
place upper limits on the Fe K$\alpha$ EWs. Only 8 of the 72 sources
($\approx$11\%) have 90\% confidence EW upper limits above 1 keV, and,
of these, none has a measured photon index $\Gamma < 1.0$. Since
$\sim$50\% of all known Compton thick sources have $\Gamma < 1.0$ and
large EWs can also arise from anisotropic ionizing radiation or
variability lags between continuum and line emissions (e.g., Maiolino
et al. 1998; Bassani et al. 1999), the true number of Compton thick
sources among our spectroscopically identified sources may be
even smaller.

\vspace{-0.12in}
\section{Variability}
\vspace{-0.11in}

The 20 individual CDF-N observations span approximately 27 months
and offer an unprecedented probe of long-term X-ray variability in
distant AGN. Using both Kolmogorov-Smirnov and $\chi^{2}$ statistics, we
find that $\sim$94\% of the BLAGN, $\sim$90\% of the $>$500 count
sources, and $\sim$60\% of the overall sample show indications of
variability. 

\begin{figure}
\vspace{-0.52in}
\resizebox{\hsize}{!}
{\includegraphics{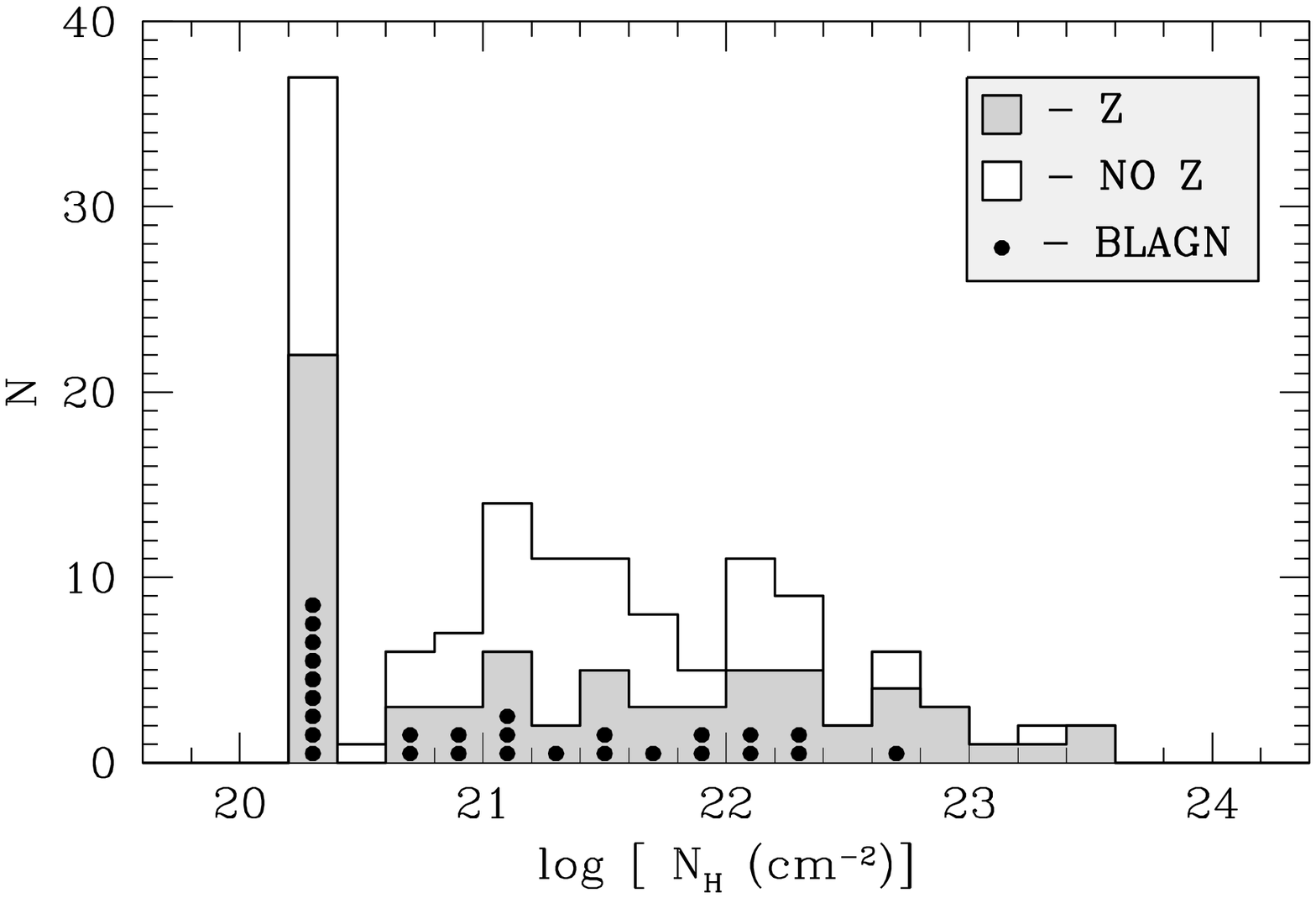}\hfill
\includegraphics{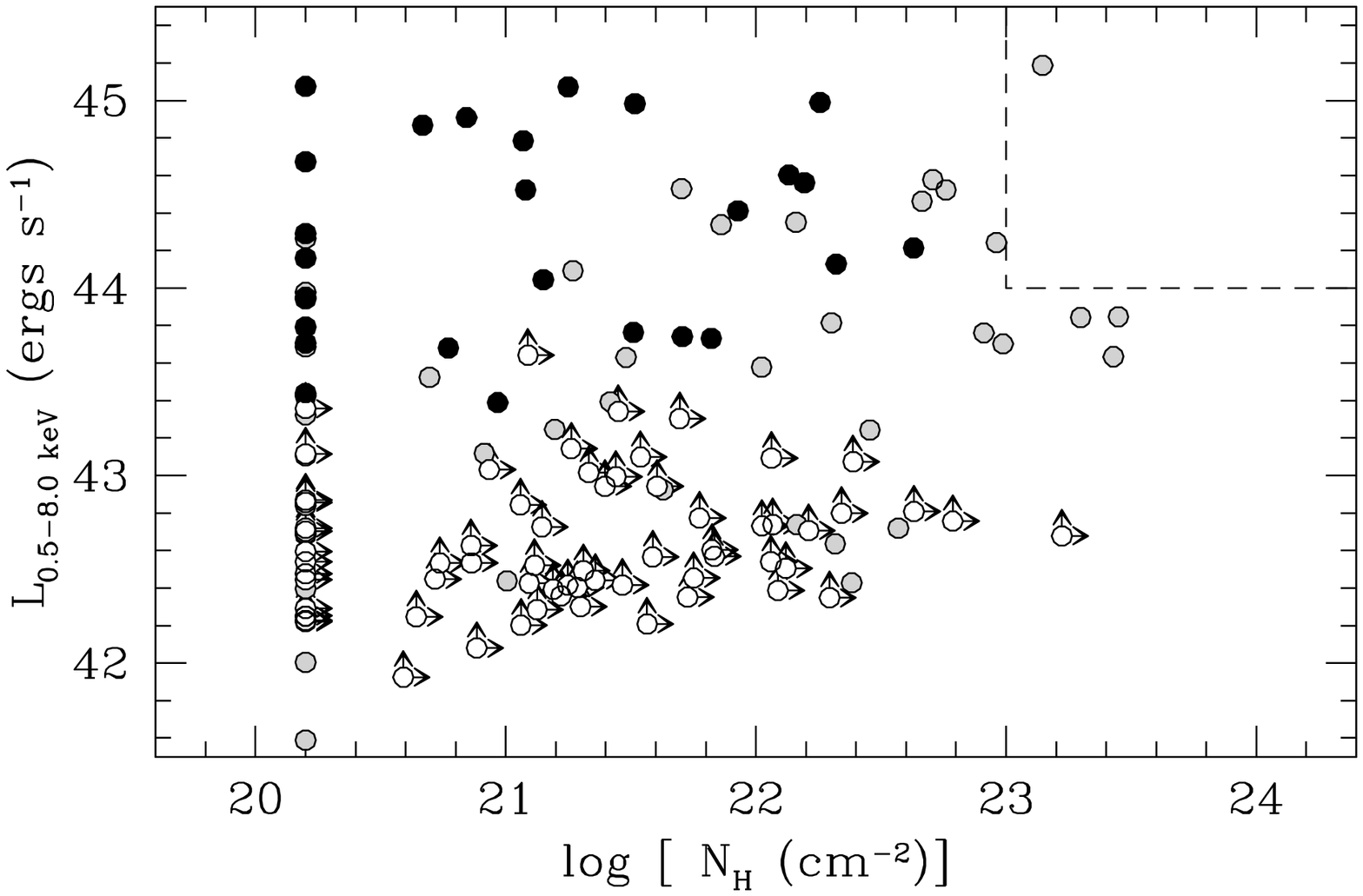}}
\caption{\small {\it Left:} The overall intrinsic $N_{\rm H}$ distribution 
for our sample. {\it Right:} $L_{\rm X}$ vs. $N_{\rm H}$ for our
sample. Sources without known redshifts are plotted as $L_{\rm X}$
lower limits assuming $z=0.5$.}
\label{fig1}
\vspace{-0.16in}
\end{figure}

\vspace{-0.12in}
\section*{Acknowledgments}
\vspace{-0.11in}

We gratefully acknowledge the financial support of NSF CAREER award
AST-9983783 (FEB, CV, DMA, WNB), NASA grant NAS 8-38252 (GPG, PI), 
NASA GSRP grant NGT5-50247 (AEH), and NSF grant AST-9900703 (DPS).

\end{document}